# Artificial intelligence applications in Parkinson's disease via retinal imaging

Ali Jafarizadeh [1, *, #], Hamidreza Ashayeri [2, #], Hadi Vahedi [1], Parsa Khalafi [3], Mirsaeed Abdollahi [1], Navid Sobhi [4], Ru-San Tan [5, 6], Roohallah Alizadehsani [7, *], U. Rajendra Acharya [8, 9]

1. Nikookari Eye Center, Tabriz University of Medical Sciences, Tabriz, Iran.
2. Research Center for Evidence-based Medicine, Iranian EBM Center: A Joana-affiliated group, Tabriz University of Medicine Science, Tabriz, Iran
3. School of Medicine, Tehran University of Medical Sciences, Tehran, Iran
4. Translational Ophthalmology Research Center, Farabi Eye Hospital, Tehran University of Medical Science, Tehran, Iran.
5. National Heart Centre Singapore, Singapore.
6. Duke-NUS Medical School, Singapore.
7. Institute for Intelligent Systems Research and Innovation (IISRI), Deakin University, VIC 3216, Australia
8. School of Mathematics, Physics and Computing, University of Southern Queensland, Springfield, Australia
9. Centre for Health Research, University of Southern Queensland, Australia

***Corresponding Authors**:

1- Roohallah Alizadehsani, PhD

*Institute for Intelligent Systems Research and Innovation (IISRI), Deakin University, VIC 3216, Australia*

*Tel: +61 3 524 79394*

*Postal: 75 Pigdons Rd, Waurn Ponds VIC 3216, Australia*

*Email:* r.alizadehsani@deakin.edu.au

https://orcid.org/0000-0003-0898-5054

2- Ali Jafarizadeh, MD, MPH

*Nikookari Eye Center, Tabriz University of Medical Sciences, Tabriz, Iran*

*Tel: +98 901 098 0062*

*Postal code:* 51666/14766

*Email:* Jafarizadeha@tbzmed.ac.ir, Ali.jafarizadeh.md@gmail.com

https://orcid.org/0000-0003-4922-1923

Author details:

| Name | Degree | Email | ORCiD |
|---|---|---|---|
| Ali Jafarizadeh | MD, MPH | ali.jafarizadeh.md@gmail.com | https://orcid.org/0000-0003-4922-1923 |
| Hamidreza Ashayeri | MD | hashayeri97@gmail.com | https://orcid.org/0000-0002-2593-9810 |
| Hadi Vahedi | MD | Hadivahedimrg@gmail.com | https://orcid.org/0009-0008-4769-451X |
| Parsa Khalafi | MD, MSc | p-khalafi@alumnus.tums.ac.ir | https://orcid.org/0009-0005-1975-9855 |
| Mirsaeed Abdollahi | MD | abdollahi.mirsaeed@gmail.com | https://orcid.org/0000-0002-2444-1172 |
| Navid Sobhi | MD | navidsbg.ns1998@gmail.com | https://orcid.org/0000-0003-0663-850X |
| Ru-San tan | MBBS | tanrsnhc@gmail.com.sg | https://orcid.org/0000-0003-2086-6517 |
| Roohallah Alizadehsani | PhD | r.alizadehsani@deakin.edu.au | https://orcid.org/0000-0003-0898-5054 |
| U. Rajendra Acharya | PhD, DEng, DSc | Rajendra.Acharya@unisq.edu.au | https://orcid.org/0000-0003-2689-8552 |

## Abstract

Parkinson's disease (PD) is projected to increase substantially due to population aging, making early diagnosis increasingly important, as timely detection may delay progression and reduce long-term complications. Retinal microvasculature has emerged as a promising anatomical biomarker of neurodegeneration, and when combined with artificial intelligence AI, retinal imaging may provide an advanced, noninvasive, and cost-effective screening strategy for PD. This study evaluated the evidence from the past 35 years regarding the capability of AI to detect early PD-related changes in retinal vascular structure. Five electronic databases including PubMed, Web of Science, Scopus, ScienceDirect, and ProQuest were systematically searched from January 1990 to January 2025. In addition, Annals of Neurology and Frontiers in Neuroscience were hand-searched, and the reference lists of included studies were screened for additional eligible publications. Nineteen studies met the inclusion criteria. Three principal diagnostic AI tasks were identified, including disease classification, retinal vessel segmentation, and PD risk stratification. The best-performing models were ShAMBi-LSTM on the Drishti dataset with 97.2 percent accuracy, 99.5 percent precision, 96.9 percent sensitivity, and an F1 score of 0.981 for classification, nnU-Net with 99.7 percent accuracy, 98.7 percent precision, 98.9 percent sensitivity, 99.8 percent specificity, and a Dice score of 98.9 percent for segmentation, and AlexNet for risk prediction with area under the curve values of 0.77, 0.68, and 0.73 across datasets. Overall, application of AI algorithms to retinal vasculature for detecting early signs of PD and predicting disease severity suggests that integration of AI with retinal biomarkers holds substantial potential for earlier and more accurate detection compared with traditional clinical evaluation alone.

**Keywords**: Artificial intelligence, Parkinson`s disease, Retinal imaging, Microvasculature, Deep learning.

Highlights:

1. Early thinning of RNFL and GCL is a key non-invasive retinal biomarker for PD.
2. AI analysis of retinal imaging can detect early PD, segment retinal layers, and estimate disease severity.
3. AI performance depends heavily on data quality, architecture choice, and advanced methods (TL and XAI)
4. The top-performing AI algorithms in different tasks were ShAMBi-LSTM for classification, nnU-Net for segmentation, and AlexNet for risk prediction.

# 1. Introduction

Parkinson's disease (PD) is a neurodegenerative disease affecting nearly 12 million people worldwide (Luo et al., 2025). PD incidence increases with age; only 4% of cases are less than 50 years old (Van Den Eeden et al., 2003). With global population aging, the prevalence of PD is estimated to increase by 76% from 2021 to 2050 (Su et al., 2025). Early diagnosis is crucial, as timely detection and management can delay progression and reduce long-term complications. Clinical diagnosis of PD is based on detailed history and neurological examination. It is important to exclude neurological conditions that mimic PD (Kobylecki, 2020) using neuroimaging, e.g., magnetic resonance imaging (MRI) (Heim et al., 2017; Pyatigorskaya et al., 2014). Accurate diagnosis enables appropriate management, but the task is onerous and requires expertise. There is an unmet need for accessible screening tools for population-level detection of PD.

Retinal imaging has been proposed as a cost-effective solution for PD screening. The retina undergoes distinct structural and functional changes in PD. PD severity and duration are negatively associated with retinal ganglion cell inner plexiform (GCL-IPL) and ganglion cell complex (GCC) thicknesses (Bayhan et al., 2014; Sari et al., 2015). The close relationship between retinal and central nervous system (CNS) pathologies is implied by several studies. Simultaneously, in PD, there are numerous degenerative, dysfunctional, and structural retinal changes. PD duration is negatively correlated with retinal nerve fiber layer (RNFL) thickness, and the Unified Parkinson's Disease Rating Scale (UPDRS) is negatively correlated with light-adapted electroretinography findings (Elanwar et al., 2023). Notably, retinal imaging data are amenable to artificial intelligence (AI)-enabled analysis. Both machine learning (ML) and deep learning (DL) are increasingly being applied in medical imaging to perform various tasks, such as image segmentation, artifact detection, quality improvement, and image classification (Thrall et al., 2018). This work reviews published research on AI applications for various aspects of PD detection using retinal imaging (Figure 1).

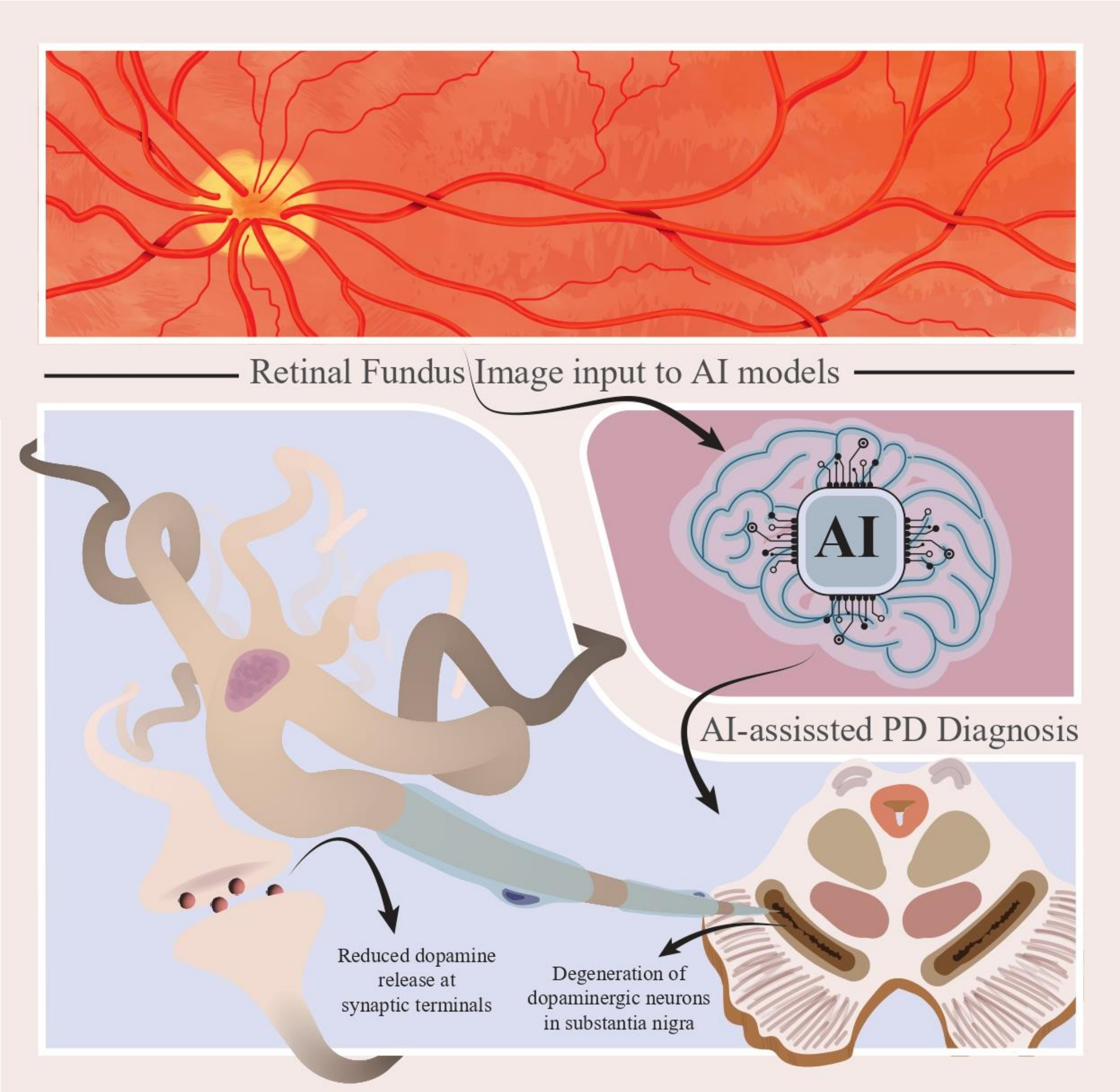

Figure 1. AI applications in Parkinson's disease detection through retinal imaging.

## 2. Method

We performed a literature search for articles published between January 1990 to January 2025 in PubMed, Web of Science (WOS), Scopus, ScienceDirect, and ProQuest. We included original research that employed ML or DL for the prediction and diagnosis of PD in patients, including image processing (e.g., image segmentation) of retinal images obtained by different methods, such as fundoscopy and optical coherence tomography (OCT). We excluded review studies, book chapters, editorials, letters, studies that did not use ML or DL, studies that did not include patients with PD, studies that did not use retinal images, and studies that did not contain at least an extended English abstract. The Boolean strings used for searching the different databases are detailed in Table S1. Using the same inclusion and exclusion criteria, we performed a hand search of articles published within the same period in two key journals, “Annals of Neurology” and “Frontiers in Neuroscience”. References of selected studies were also scanned for suitable studies. Table 1 summarizes the inclusion and exclusion criteria used to during the selection of papers.

**Table 1.** Inclusion and exclusion criteria employed to select the papers.

| Inclusion criteria | Exclusion criteria |
|---|---|
| Original studies that use:<br>• Machine learning or Deep learning algorithms and<br>• Retinal images obtained by different methods, such as fundoscopy or optical coherence tomography<br>• For prediction, diagnosis, and image segmentations<br>• In patients with Parkinson`s disease | • Review studies/ Book chapters/ Editorials/ Letters<br>• Studies that didn`t use Machine learning/Deep learning<br>• Studies that didn`t include patients with Parkinson`s disease<br>• Studies that didn`t use retinal images<br>• Studies without at least an extended English Abstract |

# 3. Results

### 3.1 Study Selection

A total of 3,323 records were identified across all sources. 1,336 duplicate studies were removed, leaving 1,987 studies. Screening of abstracts, followed by full-text assessment, results in a final selection of 19 studies (Ahn et al., 2023; Álvarez-Rodríguez et al., 2024; Chen et al., 2024; Chrysou et al., 2024; Gende et al., 2023; Hu et al., 2025; Hu et al., 2022; J. Huang et al., 2024; Lee et al., 2023; LIANG Keke, 2024; Nunes et al., 2019; Richardson et al., 2024; Rupavath et al., 2024; Shi et al., 2024; Tran et al., 2024; Ueda et al., 2024; Varghese et al., 2023; Zhang et al., 2024; Zhao et al., 2021; Zhou et al., 2023) (Figure 2). Scanning the references of selected studies did not identify any more eligible studies. In a similar study to ours, Ganji et al. (Ganji et al., 2025), aimed to assess the use of AI in detecting multiple neurodegenerative diseases based on retinal imaging. However, they have included 4 articles with PD patients and we aim to assess the 19 articles used in more extreme detail compared to Ganji et al. (Ganji et al., 2025).

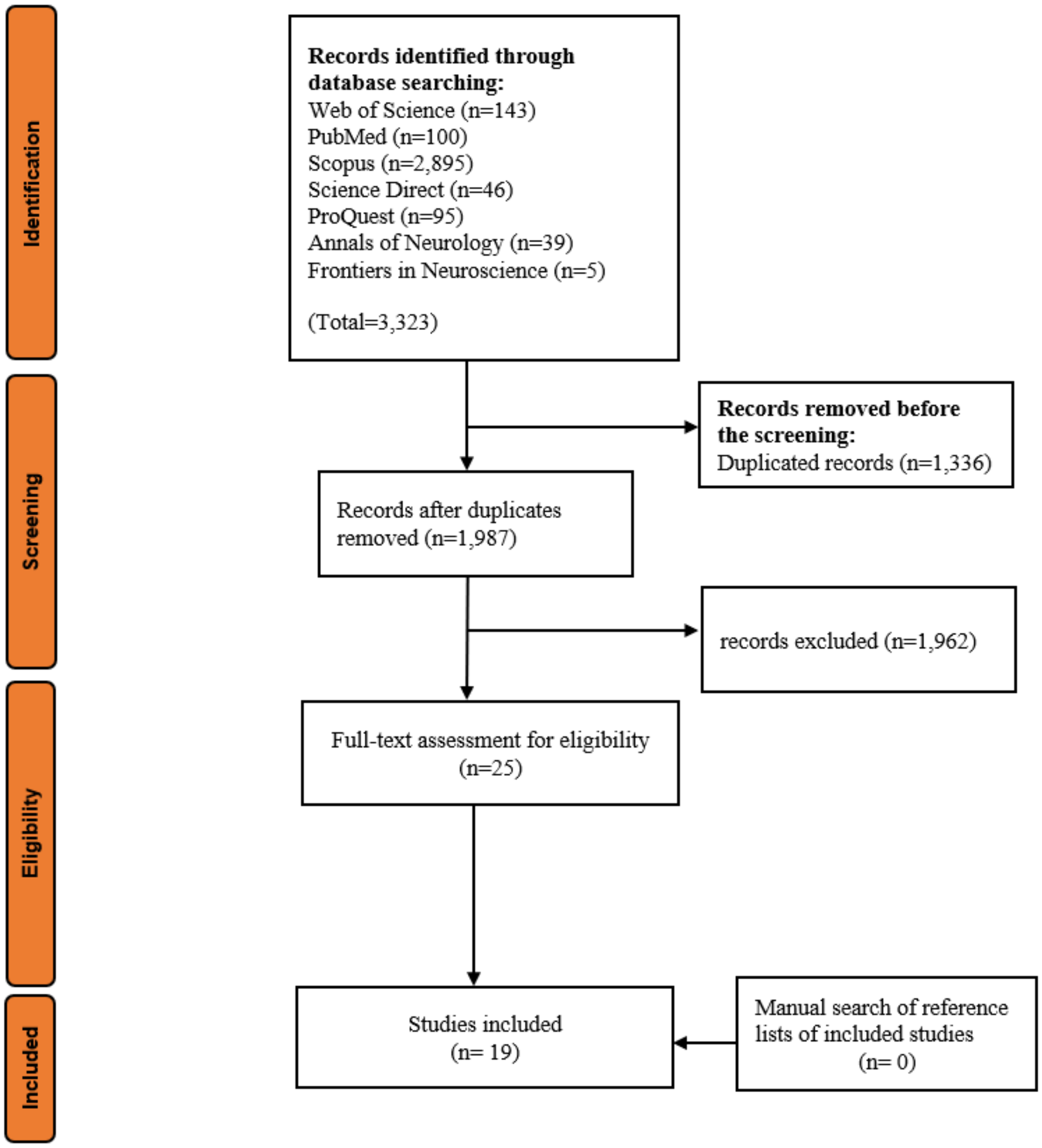

**Figure 2.** Flow diagram of study selection.

### 3.2 Study Characteristics

Table 2 summarizes the 19 selected studies. Most datasets were single-center and internally validated; only two studies (Ahn et al., 2023; Zhou et al., 2023) reported a clear external validation cohort. Several papers reported multi-class tasks involving PD patients as well as healthy controls and/or patients with other neurologic or ocular diseases (e.g., Alzheimer's disease, multiple sclerosis, glaucoma). Included studies were published during 2019-2025. Figure 3 represents the number of published articles in each year based on their approach (ML or DL). The PD sample sizes in the studies ranged from 6 (Zhang et al., 2021; Zhao et al., 2021)  to 322 (Zhou et al., 2023) patients, with most studies enrolling 20–150 PD subjects; several papers also included healthy controls and/or other neurologic or ocular diseases (e.g., AD, MS, glaucoma) to enable multi-class tasks or differential diagnosis. 9 (47.3%) studies used OCT alone (Álvarez-Rodríguez et al., 2024; Chen et al., 2024; Chrysou et al., 2024; Gende et al., 2023; Jingqi Huang et al., 2024; Nunes et al., 2019; Shi et al., 2024; Zhang et al., 2024; Zhao et al., 2021); 5 (26.3%) studies used color fundus photography (Ahn et al., 2023; Hu et al., 2025; Hu et al., 2022; Tran et al., 2024; Varghese et al., 2023); and one evaluated retinal hyperspectral imaging (Ueda et al., 2024). A few studies were multimodal (Lee et al., 2023; Richardson et al., 2024; Zhou et al., 2023), and one article didn't specify which type of retinal image they used from the online datasets (Rupavath et al., 2024).

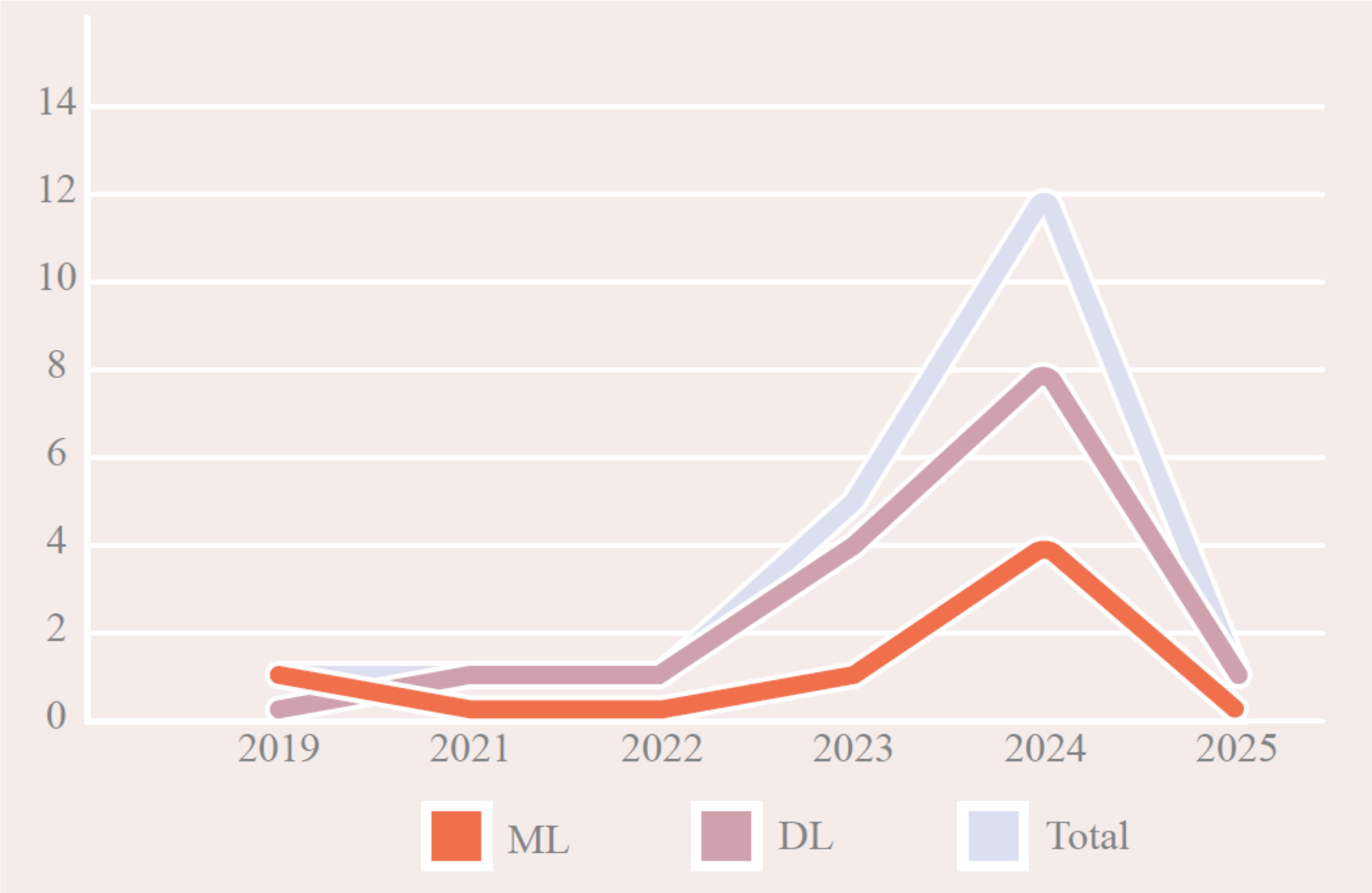

**Figure 3.** Number of published articles in each year in various categories.

**Table 2.** Summary of characteristics of included studies.

| **Author (Year)** | **Number of subjects (F/M ratio)** | **Mean Age ±Std** | **Number of used eye images (R/L ratio)** | **Retinal imaging techniques** | **Training/ Validation/ Testing (K-fold cross-validation)** | **Task** | **AI model** | **AUC** | **Accuracy/ Precision** | **Sensitivity/Specificity** | **Dice score/ F1 score** |
|---|---|---|---|---|---|---|---|---|---|---|---|
| Nunes et al. (2019) (Nunes et al., 2019) | HC: 27 (14/13)<br>PD: 28 (15/13)<br>AD: 20 (10/10) | HC: 64.1 ±7.1<br>PD: 63.4 ±6.6<br>AD: 66.3 ± 6.8 | HC: 53 (26/27)<br>PD: 54 (27/27)<br>AD: 39 (20/19) | OCT | (2, 5, 10-fold cross-validation) | Classification: HC vs PD vs AD (2-fold cross-validation) | SVM | - | 76.7% / - | HC: 84.9% / 80.6%<br>PD: 74.1% / 96.7%<br>AD:71.8% / 90.7% | - |
| | | | | | | Classification: HC vs PD vs AD (5-fold cross-validation) | | - | 80.8% / - | HC: 87.7% / 83.9%<br>PD: 75.9% / 97.8%<br>AD:79.5% / 92.5% | - |
| | | | | | | Classification: HC vs PD vs AD (10-fold cross-validation) | | - | 82.2% / - | HC: 88.7% / 84.9%<br>PD: 77.8% / 97.8%<br>AD:79.5% / 92.5% | - |
| Zhao et al. (2021) (Zhao et al., 2021) | HC: 32 (20/12)<br>PD: 6 (4/2)<br>WMH: 31 (12/19) | HC: 60.56 ± 5.72<br>PD: 62.67 ± 11.48<br>WMH: 63.58 ± 7.12 | HC: 58 (28/30)<br>PD: 11 (5/6)<br>WMH: 56 (28/28) | OCT | - | Segmentation | CE-Net (Gu et al., 2019) | - | - | - | - |
| Hu et al. (2022) (Hu et al., 2022) | 46886 (female to male ratio of 35834 PD free used for analysis | Mean age of 35834 patients used for analysis: 56.7 ± 8.04 | 19200 fundus images were used | CFP | 11052/-/35834 | Prediction (not with AI) | Xception model provided by Zhu et al. (2022) (Zhu et al., 2023) | 0.708 | - | - | - |
| | 55.7% / 44.3%) | | for age prediction | | | | | | | | |
| Ahn et al. (2023) (Ahn et al., 2023) | non-PD atypical motor abnormalities: 349 (236/ 113)<br>PD: 266 (132/ 134) | non-PD atypical motor abnormalities:<br>70.7 ±7.9<br>PD: 70.8 ± 8.3 | non-PD atypical motor abnormalities: 700<br>PD: 539 | CFP | Data were randomly obtained from 80<br>people for the validation data set, and the rest for the training<br>data set. | Classification: PD vs non-PD atypical motor abnormalities | ResNet-18 | 0.76 | 76.75% / - | 77.05%/ 76.46% | - |
| | | | | | | | ResNet-101 | 0.71 | 71.18% / - | 71.65% / 70.73% | - |
| | | | | | | | VGG19 | 0.66 | 66.30% / - | 65.61% / 66.96% | - |
| | | | | | | | EfficientNet-b0 | 0.73 | 73.57% / - | 75.11% / 72.05% | - |
| | | | | | | | EfficientNEt-b7 | 0.68 | 68.36% / - | 69.04% / 67.69% | - |
| | | | | | Model 1: fundus only. | Classification: H-Y severity (3 classes) | ResNet-18 | 0.66 | 58.24% / - | 76.76% / 52.72% | - |
| | | | | | Model 2: fundus with sex and age data. | | | 0.71 | 64.80% / - | 79.48% / 58.81% | - |
| | | | | | Model 3: fundus with sex, age, diabetes, and hypertension data. | | | 0.72 | 63.55% / - | 79.52% / 59.44% | - |
| | | | | | Model 4: fundus image with sex, age, diabetes, and hypertension data with multimodal method. | | | 0.77 | 73.38% / - | 83.23% / 66.81% | - |
| | | | | | Model 1: fundus only. | Classification: Unified Parkinson's Disease Rating Scale (UPDRS-III) (3 classes) | ResNet-18 | 0.67 | 59.43% / - | 77.16% / 54.75% | - |
| | | | | | Model 2: fundus with sex and age data. | | | 0.69 | 62.24% / - | 78.93% / 57.67% | - |
| | | | | | Model 3: fundus with sex, age, diabetes, and hypertension data. | | | 0.70 | 65.31% / - | 79.54 % / 59.21% | - |
| | | | | | Model 4: fundus image with sex, age, diabetes, and | | | 0.77 | 71.64% / - | 82.61% / 65.75% | - |
| | | | | | hypertension data with multimodal method. | | | | | | |
| Gende et al. (2023) (Gende et al., 2023) | Individuals recruited from clinics:<br>HC: 10<br>PD: 10<br>AD: 10<br>Relapsing-remitting Multiple Sclerosis: 10<br>Essential tremor: 10 | - | 732 OCT images from Li et al.`s dataset (Li et al., 2021)<br>1250 retinal images from patients recruited from clinics | OCT | The model was pretrained on 732 OCT images, and in experiment 1, it was trained on four groups (40 patients) and tested on the remaining group (10 patients) | Segmentation (RNFL / trained on whole dataset) | MGU-Net (Li et al., 2021) | - | - / 94.2% | 95.7% / - | 94.9% / - |
| | | | | | | Segmentation (GCL-BM / trained on whole dataset) | | | - / 98.4% | 98.6% / - | 98.7% / - |
| | | | | | | Segmentation (RNFL / trained on whole dataset except PD) | | - | - / 93.8% | 95.8% / - | 94.8%/ - |
| | | | | | | Segmentation (GCL-BM / trained on whole dataset except PD) | | | - / 98.4% | 98.7% / - | 98.5%/ - |
| | | | | | | Segmentation (RNFL / trained on HC dataset) | | - | - / 93% | 95.8% / - | 94.3%/ - |
| | | | | | | Segmentation (RNFL / trained on AD dataset) | | - | - / 94.6% | 93.8% / - | 94.2%/ - |
| | | | | | | Segmentation (RNFL / trained on MS dataset) | | - | - / 93.5% | 94.1% / - | 93.8%/ - |
| | | | | | | Segmentation (RNFL / trained on ET dataset) | | - | - / 93.9% | 94.4% / - | 94.1%/ - |
| | | | | | | Segmentation (GCL-BM / trained on HC dataset) | | - | - / 98.7% | 98.3% / - | 98.5%/ - |
| | | | | | | Segmentation (GCL-BM / trained on AD dataset) | | - | - / 98.2% | 98.6% / - | 98.4%/ - |
| | | | | | | Segmentation (GCL-BM / trained on MS dataset) | | - | - / 98.3% | 98.4% / - | 98.4%/ - |
| | | | | | | Segmentation (GCL-BM / trained on ET dataset) | | - | - / 98.9% | 98.5% / - | 98.7%/ - |
| Lee et al. (2023) (Lee et al., 2023) | PD: 107<br>Normal cognition: 298<br>Other diseases: 237 | - | OCT: 1465 | OCT, OCTA | 70%/15%/15% | Classification: Good vs poor image quality (GC-IPL OCT scan) | AlexNet | 0.99 | 97.1% / - | 88.9% / 97% | - |
| | PD: 107<br>Normal cognition: 425<br>Other diseases: 239 | | OCTA: 2689 | | | Classification: Good vs poor image quality (OCTA scan) | AlexNet | 0.832 | 76.4% / - | 65.7% /80.9% | - |
| Varghese et al. (2023) (Varghese et al., 2023) | HC: 528<br>PD: 133<br>AD: 18<br>Vascular dementia:528 | - | Only left eye images were used. | CFP | 60%/-/40% | | Inductive logic programming | - | - | - | - |
| Zhou et al. (2023) (Zhou et al., 2023) | MEHAlzEye:<br>1) Control: 293 (144/149) | - | - | CFP, OCT | 904170 CFP images and 736442 OCT were used for construction, MEHAlzEye data were used for fine-tuning and internal | Prediction: Incident PD vs non-incident PD | RETFound | 0.669 | - | - | - |
| | 2: PD: 293 (120/173)<br>UK Biobank:<br>1) Control: 29 (12/17)<br>2) PD: 29 (19/10) | | | | validation, and UK Biobank was used for external validation | | | | | | |
| Álvarez-Rodríguez (2024) et al. (Álvarez-Rodríguez et al., 2024) | HC: 81<br>PD: 82<br>AD: 29<br>Multiple Sclerosis: 166<br>Essential tremor: 10 | - | - | Single-view and Multiview OCT | 70%/10%/20% | Segmentation: RNFL | nnU-Net architecture (Macula view-2D approach) | - | 99.8% / 93.7% | 96.3% / 99.9% | 95.5% / - |
| | | | | | | Segmentation: GCL - Bruch's Membrane | | | 99.7% / 98.7% | 98.7% / 99.8% | 98.8% / - |
| | | | | | | Segmentation: RNFL | nnU-Net architecture (Macula view-3D multi-scale cascade approach) | - | 99.7% / 92% | 93.4% / 99.9% | 94.1% / - |
| | | | | | | Segmentation: RNFL | | | 99.6% / 98.4% | 98.5% / 98.6% | 98.5% / - |
| | | | | | | Segmentation: RNFL | nnU-Net architecture (Macula view-3D full resolution approach) | - | 99.8% / 94.6% | 95.8% /99.9% | 95.7% / - |
| | | | | | | Segmentation: GCL - Bruch's Membrane | | | 99.7% / 98.7% | 98.9% / 99.8% | 98.9% / - |
| | | | | | | Segmentation: RNFL | nnU-Net architecture (Optic view) | - | 98.9% / 90.6% | 89.9% / 99.5% | 90% / - |
| | | | | | | Segmentation: GCL - Bruch's Membrane | nnU-Net architecture (Optic view) | - | 97.9% / 93.4% | 85.5% / 99.3% | 88.6% / - |
| | | | | | | Segmentation: RNFL | Transfer learning (Optic view) | - | 99.1% / 91.7% | 92.0% / 99.5% | 91.6% / - |
| | | | | | | Segmentation: GCL - Bruch's Membrane | Transfer learning (Optic view) | - | 98.2% / 92.1% | 91.6% / 99.0% | 91.2 % / - |
| Chen et al. (2024) (Chen et al., 2024) | PD: 45 | - | 630 | OCT | 441/-/189 | Segmentation of retinal layers | Wavelet Transformer (based on U-net architecture) | - | 93.73% /- | - | 93.46% / - |
| Chrysou et al. (2024) (Chrysou et al., 2024) | HC: 100 (54/46)<br>PD: 121 (34/87)<br>Primary open-angle Glaucoma: 78 (45/33) | Median age of (IQR)<br>HC: 62.7 (6.5)<br>PD: 65.1 (8.7)<br>Primary open-angle Glaucoma: 65.7 (9.7) | - | SD-OCT | 141/36/44 | Classification: PD vs HC | RF | 0.71 | 57% /- | - | - |
| | | | | | 128/32/39 | Classification: PD vs primary open-angle glaucoma | RF | 0.92 | 82% / - | - | - |
| Huang et al. (2024) (J. Huang et al., 2024) | HC: 49 (32/17)<br>PD: 26 (11/15) | HC: 61.61 ± 5.94<br>PD: 65.73 ± 6.90 | HC: 1584<br>PD: 666 | OCT | 60%/20%/20% (five-fold cross-validation) | Classification: PD vs HC | Wavelet-based Selection and Recalibration Network (WaveSRNet) | 0.7516 | 74.49%/ - | - | - / 0.5676 |
| Keke et al. (2024)* (LIANG Keke, 2024) | PD: 49<br>HC: 39 | PD: 40-79 | - | OCT, OCTA | 62 training/26 validation | Classification: PD vs HC | LR | 0.841 | -/- | - | - |
| | | | | | | | Decision tree | 0.806 | -/- | - | - |
| | | | | | | | RF | 0.766 | -/- | - | - |
| | | | | | | | KNN | 0.764 | -/- | - | - |
| | | | | | | | XGBoost | 0.732 | -/- | - | - |
| Richardson et al. (2024) (Richardson et al., 2024) | Control: 249 (184/67)<br>PD: 52 (21/31) | Control: 66.4 ± 9.1<br>PD: 65.9 ± 9.5 | Control: 371 (204/169)<br>PD: 75 (39/36) | OCT, OCTA, UWF color photographs, UWF autofluorescence | 355 eyes/ 42 eyes/ 49 eyes | Classification: PD vs HC | CNN (OCTA, GC-IPL, UWF FAF, UWF color - Right eye only) | 0.91 | - | 100% / 85% | - |
| | | | | | | | CNN (OCTA, GC-IPL, UWF FAF, UWF color - Left eye only) | 0.918 | - | 100% / 85.7% | - |
| | | | | | | | CNN (OCTA, GC-IPL, UWF FAF, UWF color - Both eyes) | 0.861 | - | - | - |
| | | | | | | | CNN (OCTA - Right eye only) | 0.711 | - | - | - |
| | | | | | | | CNN (OCTA - Left eye only) | 0.811 | - | - | - |
| | | | | | | | CNN (OCTA - Both eyes) | 0.650 | - | - | - |
| | | | | | | | CNN (GC-IPL - Right eye only) | 0.630 | - | - | - |
| | | | | | | | CNN (GC-IPL - Left eye only) | 0.556 | - | - | - |
| | | | | | | | CNN (GC-IPL - Both eyes) | 0.625 | - | - | - |
| | | | | | | | CNN (UWF FAF - Right eye only) | 0.682 | - | - | - |
| | | | | | | | CNN (UWF FAF - Left eye only) | 0.794 | - | - | - |
| | | | | | | | CNN (UWF FAF - Both eyes) | 0.855 | - | - | - |
| | | | | | | | CNN (UWF color - Right eye only) | 0.798 | - | - | - |
| | | | | | | | CNN (UWF color - Left eye only) | 0.838 | - | - | - |
| | | | | | | | CNN (UWF color - Both eyes) | 0.805 | - | - | - |
| Shi et al. (2024) (Shi et al., 2024) | HC: 49<br>PD: 26 | - | HC: 1584<br>PD: 666 | OCT | Eye-wise five-fold cross-validation | Classification: PD vs HC (Eye-wise) | RSGA-Net | - | 88.21% / 81.28% | 81.67% / 90.97% | - / 0.8068 |
| | | | | | | | RSGA-Net-V1 | - | 87.95% / 83.51% | 75.08% / 93.40% | - / 0.7842 |
| | | | | | subject-wise five-fold cross-validation | Classification: PD vs HC (Subject-wise) | RSGA-Net | - | 83.38% / 61.81% | 21.39% / 96.43% | - / 0.2899 |
| | | | | | | | RSGA-Net-V1 | - | 79.95% / 49.98% | 29.17% / 90.64% | - / 0.307 |
| Tran et al. (2024) (Tran et al., 2024) | HC: 84 (35/49)<br>PD: 84 (35/49) | HC: 61.8 ± 5.9<br>PD: 61.8 ± 5.9 | HC: 123<br>PD: 123 | CFP | five-fold stratified cross-validation (3 folds for training, one for validation, and one for testing) | Classification: PD vs HC | LR | 0.68 | 63%/ - | 66% / 61% | - / 0.64 |
| | | | | | | | Elastic Net | 0.67 | 61%/ - | 68% / 54% | - / 0.64 |
| | | | | | | | SVM (linear) | 0.69 | 63%/ - | 65% / 62% | - / 0.64 |
| | | | | | | | SVM (RBF) | 0.71 | 67%/ - | 76% / 58% | - / 0.7 |
| | | | | | | | AlexNet | 0.77 | 68%/ - | 76% / 60% | - / 0.68 |

| | | | | | | | | | | |
|---|---|---|---|---|---|---|---|---|---|---|
| | | | | | | | VGG-16 | 0.71 | 66%/ - | 85% / 46% | - / 0.7 |
| | | | | | | | GoogleNet | 0.69 | 63% / - | 77% / 49% | - / 0.67 |
| | | | | | | | Inception-V3 | 0.56 | 54% / - | 68% / 40% | -/ 0.57 |
| | | | | | | | ResNet-50 | 0.57 | 54% / - | 52% / 55% | -/ 0.45 |
| Zhang et al. (2024) (Zhang et al., 2024) | HC: 31 (20/11)<br>PD: 6 (4/2)<br>WMH: 28 (12/16) | HC: 59.1 ± 6.3<br>PD: 62.7 ± 11.5<br>WMH: 64±7.5 | HC: 57 (27/30)<br>PD: 11 (5/6)<br>WMH: 51 (25/26) | OCT | The model was trained based on the data of all four devices and then tested on the data of three of the datasets individually. | Segmentation: all retinal layers (Heidelberg Spectralis) | SGA | - | - | - | 94.43%/ - |
| | | | | | | Segmentation: all retinal layers (Optovue RTVue) | SGA | - | - | - | 87.67%/ - |
| | | | | | | Segmentation: all retinal layers (Zeiss Cirrus HD-OCT 5000) | SGA | - | - | - | 82.48%/ - |
| Ueda et al. (2024) (Ueda et al., 2024) | Control: 20 (10/10)<br>PD: 20 (7/13) | Control: 70.2 ± 10.5<br>PD: 71.0 ± 8.4 | Control: 20 (16/4)<br>PD: 20 (15/5) | Retinal hyperspectral imaging | leave-one-out cross-validation (LOOCV) | Classification: PD vs HC (Supra-nasal region) | Linear discriminant analysis | 0.6 | 57.5% / - | 60% / 55% | - |
| | | | | | | Classification: PD vs HC (Inferonasal region) | | 0.6 | 55% / - | 60% / 50% | - |
| Rupavath et al. (2024) (Rupavath et al., 2024) | - | - | - | retinal images from RIM-One and Drishti (Not-specified) | - | Classification: Neurodegenerative disorders (Not specified / RIM-One dataset) | ShAMBi-LSTM | - | 96.5% / 99.2% | 95.3% / - | - / 0.978 |
| | | | | | | Classification: Neurodegenerative disorders (Not specified / Drishti dataset) | ShAMBi-LSTM | - | 97.2% / 99.5% | 96.9% / - | - / 0.981 |
| Hu et al. (2025) (Hu et al., 2025) | PD: 263 | - | - | CFP | - | Classification: PD, MS, AMD, Glaucoma | FundusNet<br><br>(CNN and vision transformer-based) | 0.75 | - | - | - |

*The extended abstract of this article, written in English, was included in our study. However, the accuracy, precision, sensitivity, and specificity were reported in the main text study, but they were written in Chinese and, for this reason, were not included in the report. **AD:** Alzheimer`s Disease, **AMD**: Age related macular degeneration, **CFP:** Color Fundus Photography, **CNN:** Convolutional Neural Network, **GCL:** Ganglion Cell Layer, **HC:** Healthy Control, **KNN:** K-Nearest Neighbor, **LR:** Logistic Regression, **OCT:** Optic Coherence Tomography, **OCTA:** Optic Coherence Tomography Angiography, **PD:** Parkinson`s disease, **RNFL:** Retinal Nerve Fiber Layer, **RF:** Random Forest, **SVM:** Support Vector Machine, **UWF:** Ultra-Wide Field, **WMH:** White Matter Hyperintensity, **ShAMBi-LSTM:** Shuffle Attention Mechanism depends on Bi-directional Long Short-Term Memory, **MS**: Multiple sclerosis, , **SGA**: selective-guided adversarial network

The majority were classification studies (e.g., PD vs HC; PD vs atypical motor disorders; severity grading such as Hoehn–Yahr scale or UPDRS-III) (Ahn et al., 2023; Álvarez-Rodríguez et al., 2024; Chrysou et al., 2024; Hu et al., 2025; Jingqi Huang et al., 2024; Lee et al., 2023; Richardson et al., 2024; Rupavath et al., 2024; Shi et al., 2024; Tran et al., 2024; Ueda et al., 2024), while others focused on retinal layer segmentation (RNFL and/or GCL-BM, full-stack retinal segmentation) (Álvarez-Rodríguez et al., 2024; Chen et al., 2024; Gende et al., 2023; Zhang et al., 2024; Zhao et al., 2021) and PD prediction (Hu et al., 2022; Tran et al., 2024; Zhou et al., 2023). Across classification papers, AUC ranged from 0.6-0.918, with a multimodal fundus/OCTA/Ultrawide field retinal imaging (UWF) model reaching the highest accuracy; segmentation studies on macular OCT commonly achieved Dice 0.88–0.99 for GCL–BM, 0.90–0.96 for RNFL, and full-retina Dice ≈ 0.82–0.94.

# 4. Discussion

## 4.1 Classification

A support vector machine (SVM) is a ML algorithm used to identify hyperplanes and categorize data into distinct groups (Cosma et al., 2017). SVM is mainly classified as a supervised learning model and can perform linear and non-linear classifications (An et al., 2021; Mack, 2014; Winters-Hilt & Merat, 2007). Nunes et al. (2019) (Nunes et al., 2019) used multiple non-linear SVM models to differentiate between healthy controls (HC), AD, and PD. They included OCT images from a local dataset comprising 75 participants. The gray-level co-occurrence matrix was used for the feature extractor. In the next stage, each SVM model was used as a binary classification between two of the three possible groups. Each model also assessed each of the five retinal layers individually to make the diagnosis. They claim that excluding the RNFL layer improved classification performance, and it was excluded. The best reported accuracy was achieved with a five-fold cross-validation training model, ranging from 75.3% to 87.7% (median = 80.8%). They also report that 94.4% (median; range: 88.9-96.6%; k=10) of diagnoses were correct when both eyes of a participant received the same classification. In all scenarios, the model achieved the highest specificity for PD and the highest sensitivity for HC, compared with AD. This finding may be due to the larger sample sizes used for PD and HC compared with AD.

Temporal RNFL layer thickness correlates with functional PD scoring (Chang et al., 2022; Shafiee et al., 2019). The correlation between retinal thickness and PD severity may be due to dopamine's role in maintaining retinal thickness (Ahn et al., 2018; Svetel et al., 2025). Ahn et al. (2023) (Ahn et al., 2023) developed five CNN models to differentiate PD from patients with atypical motor abnormalities. ResNet-18 achieved the best results with an area under the curve (AUC) of 0.76 and an accuracy of 76.75%. Then, they used the ResNet-18 model as a basis to predict H-Y scale and UPDRS-III (PD severity assessment) scores from color fundus photography (CFP) and demographic data, including age, sex, hypertension status, and diabetes status. For the H-Y stage (3-class), the best model (Model 4) achieved a sensitivity of 83.23%, a specificity of 66.81%, an AUROC of 0.77, and an accuracy of 73.38% on internal data. For UPDRS-III (3-class), Model 4 reached a sensitivity of 82.61% (81.38–83.83), a specificity of 65.75%, an AUROC of 0.77, and an accuracy of 71.64% on internal data. In the external H–Y test set, performance was sensitivity 70.73%, specificity 66.66%, AUROC 0.67, and accuracy 70.45%.

Álvarez-Rodríguez (2024) (Álvarez-Rodríguez et al., 2024) evaluated screening performance across single-view (optic disc or macula) and multi-view (fusion of both) configurations, using F1-score as the primary metric due to class imbalance for classification of PD from other neurological diseases. The optic-disc single-view yields the weakest results, with PD F1 of approximately 0.50–0.6. Macular 2D representations improved slightly. The strongest single-view performance is achieved by macular 3D volume features, with PD F1 typically 0.70–0.80. Multi-view 2D (macula 2D + disc) sits around 0.60–0.70, showing a gain over either 2D view alone but still below macula-3D. Multi-view 3D and 2D+3D fusion generally matches macula-3D at ~0.70–0.80 with slightly lower variance rather than a consistent jump. In short, the ranking is: macula-3D ≳ multi-view (3D or 2D+3D) > multi-view 2D > macula-2D >> optic disc only. An important innovation of this study is the combination of nnU-Net for segmentation, the DieT feature extractor, and a histogram-based gradient-boosting classification tree for classification. Although they don't compare their results with other combinations, properly concatenating multiple methods can improve performance.

Lee et al. (2023) (Lee et al., 2023) utilized the AlexNet model to classify high- and low-quality images in OCT and optical coherence tomography angiography (OCTA) images. The model trained on OCT images achieved an AUC of 0.99. The superficial capillary plexus, identified from OCTA images, was also used as an independent input to AlexNet, yielding an AUC of 0.832. The authors suggested that the better AI performance in OCT may be due to the more accurate color maps used in OCT.

AI can detect and learn specific patterns in data and use them for classification. For example, PD and primary open-angle glaucoma (POAG) both result in retinal thinning (UlAin et al., 2025). Chrysou et al. (2024) (Chrysou et al., 2024) found that retinal layers, except retinal pigmented epithelium (RPE), were thinner in POAG compared to PD. The difference was most significant in GCL thickness (26.4 μm vs. 34.8 μm), and the RPE thickness difference between POAG and PD was lowest (14.2 μm vs. 14 μm). Also, when PD was compared to HC, the retina was thinner, and the difference was the greatest in the inner plexiform layer (IPL) (34.2 μm vs. 36.1 μm) and the outer plexiform layer (OPL) (22.3 μm vs. 23.1 μm). Subsequently, they trained a random forest (RF) model to distinguish between PD and POAG and between PD and HC. RF randomly selects a feature from the data and then grows decision trees. The classification is done based on the results of all grown decision trees (Salman et al., 2024). Chrysou et al.'s RF model successfully identified the former layers (GCL, RPE, IPL, and OPL) as the best features in differentiating PD from POAG and HC. The AUROC of the model in differentiating PD from POAG was higher than that in differentiating PD from HC (0.92 vs. 0.71), which may be due to a greater thickness difference between PD and POAG.

Richardson et al. (2024) (Richardson et al., 2024) used VGG-19 (a CNN-based algorithm) for feature extraction from multiple retinal images, including OCT, OCTA, and UWF scanning laser ophthalmoscopy. Four feature maps were generated by VGG-19 and then subjected to modality–specific feature transformations to estimate PD probability and classify PD. The VGG-19 model was pretrained on the ImageNet database, fine-tuned for PD, and applied in this research. The model trained on the left eye performed best (AUC = 0.918), while the model trained on both eyes performed worst (AUC = 0.861). Areas under the precision-recall curves for the left eye, the right eye, and both eyes were reported to be 0.677, 0.535, and 0.405. However, the authors emphasize that the 95% CIs for these models overlapped substantially, so no statistically significant difference can be concluded for either the both-eyes vs single-eye models or the left-eye-only vs right-eye-only models. The best performance was achieved when all imaging modalities were used as inputs. The ultra-widefield color photography performed best, and the GCL-IPL map performed worst when included as a single input.

Keke et al. (2024) (LIANG  Keke, 2024) developed five different models to distinguish between PD and HC. The logistic regression model achieved the highest AUC (0.841) compared to RF, Xception, decision trees, and K-nearest neighbors. Shi et al. (2024) (Shi et al., 2024), introduced a retinal-structure-guided CNN for early PD detection that combines 2 CNN methods (U-Net and Res-Net18). A U-Net first segments 10 retinal layers on each OCT; these anatomy maps are then aligned and fused with early image features, and a structure-guided attention module learns layer-specific weights so that PD-relevant layers contribute more to the final decision. The whole system is trained end-to-end, embedding anatomical priors directly into the classifier rather than relying solely on raw pixels. They employed two different five-fold cross-validations: the eye-wise five-fold cross-validation yielded higher accuracy and precision (88.21% / 81.28%) compared to the subject-wise five-fold cross-validation (83.38% / 61.81%). Assigning a subject’s left and right eyes to different sets significantly affects PD recognition performance.

Rupavath et al. (2024) (Rupavath et al., 2024) extracted the gray-level co-occurrence matrix from retinal pictures and used it to classify early neurodegenerative disease. This study employed the ShAM-Bi-LSTM model, achieving accuracies of 96.5% and 97.2% in the RIM-One and Drishti databases, respectively. This article does not specify the types of neurodegenerative diseases included. Additionally, the databases used in this study may not be suitable for this research, as their primary focus is on non-neurodegenerative diseases, such as glaucoma, and they include a sparse set of modalities. Still, they do not specify which ones they actually used for training and testing (Chakravarty & Sivaswamy, 2017; Fumero et al., 2011; Sivaswamy et al., 2014). Hu et al. (Hu et al., 2025) used several transformer-based and CNN models and then employed ensemble methods for classification. Their combination employs Nfnet, RegNet, and EfficientNet, alongside ViT methods (VOLO, XCIT, CIAT), as feature extractors, and multiple CNN and ViT models, combined with ensemble techniques, as classifiers. This method resulted in an AUC of 0.75 for PD.

PD screening is not a routine practice in current medical care. However, with the introduction of newer treatment options, screening and prophylaxis treatment may become available in the future. WaveSRNet, introduced by Huang et al. [32], is an AI model designed for this purpose. This model aims to preserve the high-frequency features while extracting low-frequency features and reducing image noise. The RNFL, GCL-IPL, OPL, and outer nuclear layers were the main layers used by this algorithm for PD screening. This method results in an AUC of 0.7516 and an accuracy of 74.49% in PD screening. Visualization (Grad-CAM, SHAP) indicates the model attends to specific layers and fovea-centered macular regions, supporting that PD-related changes are layer- and region-dependent rather than diffuse (Veys et al., 2019).

While many models focus on using OCT, OCTA, or CFP for training AI models, Ueda et al. (2024) (Ueda et al., 2024) used data of retinal hyperspectral imaging as input for linear discriminant analysis (LDA). They hypothesized that α-synuclein accumulation in the retina can alter retinal light-scattering properties. They observed that, at short wavelengths, retinal scattering differed with HC, especially in the supra-nasal and inferonasal regions. The retinal reflectance spectra in both supra- and inferotemporal regions of PD overlapped with HC and were not used for differentiation. This method yielded an AUC of 0.6 in classifying PD and HC. Although this study found the difference between PD and HC eyes to be in the nasal section of the retina, most of the studies report that the temporal side is more affected in PD (Di Pippo et al., 2023). However, some studies still report the nasal part to be more affected due to the PD (Svetel et al., 2025). The observed overlap in retinal temporal regions may be due to the small sample size; future studies can address this issue by increasing the number of participants.

The included studies have employed various ML and DL models in this section; the most common are CNN-based algorithms. The direct comparison between models in this task is tricky. The models have incorporated data from multiple databases, employed different approaches targeting specific membranes, and compared PD across conditions. There may be substantial differences in data quality across studies, and the features of each modality and layer affect overall performance. Model performance has also been shown to vary when PD is compared to different conditions (Chrysou et al., 2024). Although some extremely high performances observed in the studies may be due to overfitting (Álvarez-Rodríguez et al., 2024; Lee et al., 2023). The CNN-based models, particularly nnU-Net (Álvarez-Rodríguez et al., 2024) and AlexNet (Lee et al., 2023), achieved the highest AUC and accuracy; however, when studies with more than 200 PD patients were considered, ResNet-18 (Ahn et al., 2023) performed best.

**4.2. Segmentation**

Context encoder network (CE-Net) is a DL algorithm proposed by Gu et al. (2019) (Gu et al., 2019) for medical image segmentations. CE-Net segments 2D images using three main components: a feature encoder, a content extractor, and a feature decoder. The encoder used in this model is a pre-trained ResNet-34 architecture trained on the ImageNet dataset. The content extractor uses a dense atrous convolutional block and a residual multi-kernel pooling block to minimize spatial data loss. Zhao et al. (2021) (Zhao et al., 2021) used the CE-Net model for OCT image segmentation and determined the retinal layer thickness changes in PD. They report that the interdigitation zone and retinal pigment epithelium-Bruch membrane complex were thicker in PD compared to healthy controls. However, their study included only six participants with PD and utilized a cross-sectional design.

The multi-scale GCN-assisted U-shape network (MGU-Net) was first developed by Li et al. (2021) (Li et al., 2021). It aimed to segment retinal layers from OCT images in different patients using a U-Net-based approach. This method was superior to similar algorithms, but it required that the input images be standardized in a specific manner. Gende et al. (2023) (Gende et al., 2023) applied the MGU-Net to retinal layer segmentation of different neurological diseases, including PD. The segmentation task is performed in two stages: first, the retina is separated from background noise, and then the layers are identified. The best PD performance was achieved by multi-disease training: leave-one-disease-out (train on other NDDs, test on PD) yielded RNFL precision of 0.938 and GCL–BM precision of 0.984; pooling all diseases produced essentially the same results. Training on a single disease and testing on PD yielded weaker results for RNFL (precision of 0.930–0.946 with lower Dice scores), while GCL–BM remained near-ceiling (~0.983–0.989 Dice). The main limitation of this study is treating GCL-BM as a single complex, as in the study by Álvarez-Rodríguez (Álvarez-Rodríguez et al., 2024).

They reported uniformly strong layer masks across views: on macula OCT, the RNFL dice score was 0.955, whereas the GCL–BM score was near the ceiling at 0.985; moving from 2D to 3D macular pipelines did not materially change these values. In the optic-disc view, RNFL dice was slightly lower (0.90), whereas GCL–BM was 0.88. Transfer learning (TL) to the disc view primarily reduced training time, yielding metrics comparable to those obtained by training from scratch.

Chen et al. (Chen et al., 2024), similar to Hu et al. (Hu et al., 2025), developed a WaveFormer that combines two primary methods for image segmentation: CNN-based and Transformer-based segmentation. They aim to build a model that can leverage both global context information via Transformer-based segmentation and local information via CNNs. Their model used a U-Net architecture, with Transformer-based segmentation during the contraction phase and convolution during the expansion phase. This method achieved the highest Dice score (93.46%), accuracy (93.73%), and Intersection-over-Union (IoU) (87.79%), and the lowest average 95% Hausdorff distance (2.63), compared with other models, such as U-Net and Att-UNet. Figure 4 shows the best-performing AI algorithms for classification and segmentation in PD detection.

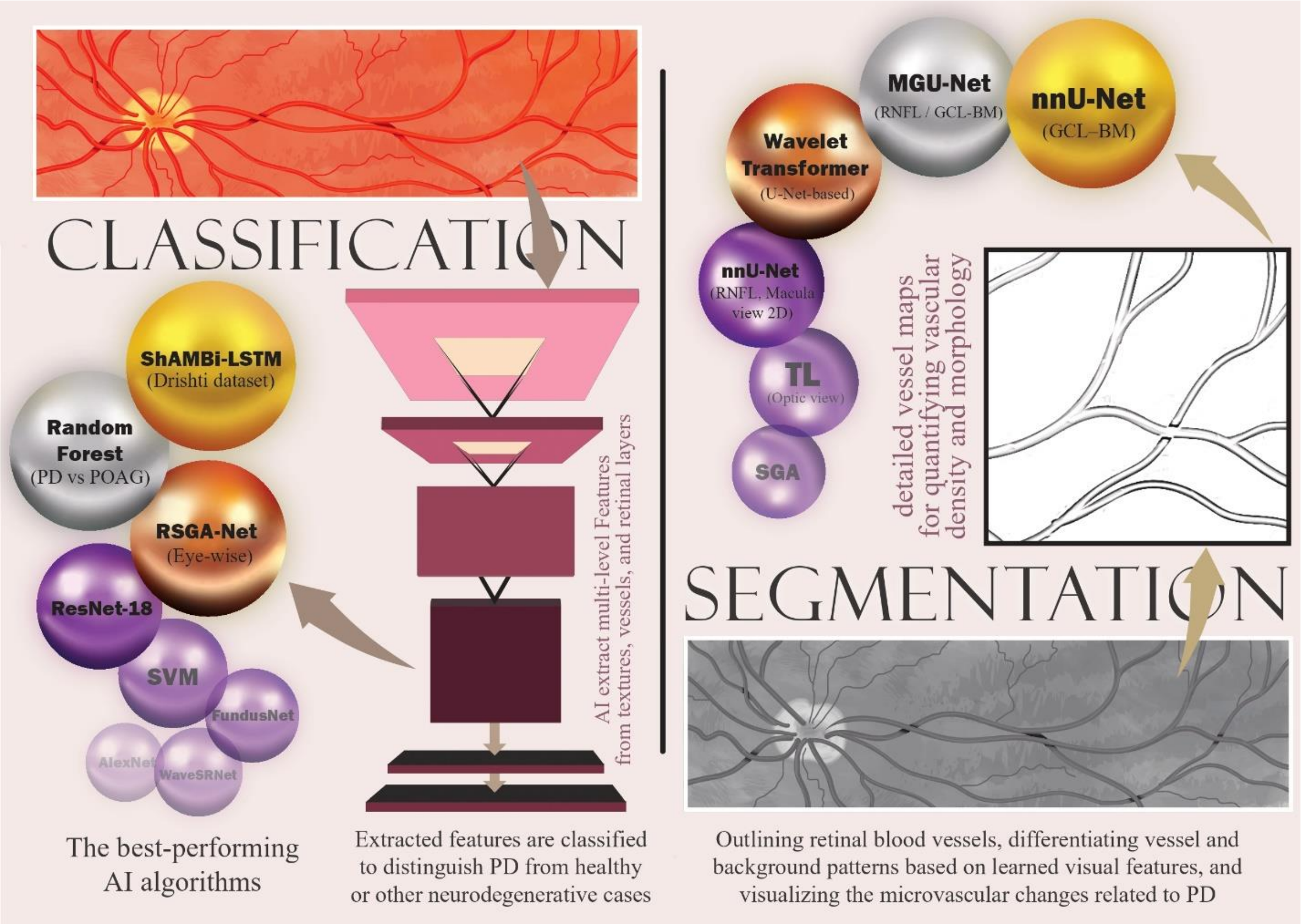

**Figure 4.** Presentation of the top-performing AI models for classification and segmentation in PD detection.

It has been demonstrated that image quality can significantly impact the performance of AI models (Gong et al., 2023). Models that perform well on one dataset often show a significant drop in performance when tested on another. In clinical practice, even the device used to capture the image can affect data quality. Zhang et al. (2024) (Zhang et al., 2024) developed a segmentation algorithm trained on four devices and tested on three. A critical aspect of their model is that it aims to segment all retinal layers, rather than segmenting them into two sections: the RNFL and GCL-BM. Their SGA model achieved an IoU of 89.5%, a dice score of 94.43%, and a mean absolute error of 17.679 in segmenting the choroid layer of the eye. For complete retinal segmentation, it performed IoU, dice score, and mean absolute error of 70.62%, 82.48%, and 5.953, respectively. Segmenting only the outer retina yielded higher performance, with an IoU of 78.23%, a dice score of 87.67%, and a mean absolute error of 4.038.

U-Net-based approaches are commonly used for image segmentation in patients with PD and perform better on OCT images, particularly in identifying the RNFL layer. The addition of the Walvelt spatial attention block or building model to a two-stage approach, such as MGU-Net, may further enhance segmentation performance. The MGU-Net may achieve the best performance in this area due to the background-removal stage precedes segmentation.

**4.3. PD Risk Prediction**

The retinal age gap, defined as the difference between deep learning predicted retinal age from fundus photographs and chronological age, has emerged as a systemic risk signal; for example, Zhu et al. (2022) (Zhu et al., 2023) showed that a positive retinal age gap is associated with higher all-cause mortality. Building on this concept, Hu et al. (2022) (Hu et al., 2022) tested whether the retinal age gap predicts incident Parkinson's disease. The underlying rationale is that retinal tissue undergoes neurodegenerative and vascular changes that parallel those in the brain; therefore, accelerated retinal ageing may signal an elevated risk of PD. In the UK Biobank cohort of 35,834 PD-free participants (median follow-up 5.83 years; 63 incident PD), each 1-year increase in retinal age gap was associated with a 10% higher risk of developing PD (hazard ratio 1.10, 95% CI 1.01–1.20; $p = 0.023$). Compared with the lowest quartile, the third quartile had a hazard ratio of 2.66 (1.13–6.22), and the highest quartile had a hazard ratio of 4.86 (1.59–14.8). As a 5-year predictor, the retinal-age model achieved an AUROC of 0.708, similar to a simple risk-factor model (0.717, not significantly different).

An article by Zhou et al. (2023) (Zhou et al., 2023) investigated the role of RETFound in PD incidence prediction based on OCT and CFP images. They included 588 participants from the MEH-AlzEye database for internal validation and 58 from the UK-Biobank for testing. They report an AUC of 0.669 for PD prediction. Their algorithm also outperformed comparable AI models (SL-ImageNet, SSL-Image-Net, SSL-Retina) when CFP images were used as input (P-value < 0.003). However, despite the higher performance, the difference was not statistically significant when OCT images were used to predict PD incidence (P = 0.085). They also observed that, in the internal dataset, RETFound outperformed CFP in predicting PD incidence from OCT images.

To predict the prevalence and incidence of PD, Tran et al. (Tran et al., 2024), developed nine different AI models. AlexNet achieved the best overall, incidence, and PD prevalence prediction performance (AUC = 0.77, 0.68, and 0.73, respectively). Other AI algorithms (Logistic regression, VGG-16, SVM, Elastic Net, Google-Net, Inception-V3, ResNet-50) achieved AUCs of 0.56-0.71, 0.56-0.73, and 0.54-0.64 in overall PD prediction, prevalence prediction, and incidence prediction, respectively. The database used in this study is relatively small, which may affect the model's performance. Although the type and quality of the data used differed between Tran et al. (Tran et al., 2024), and Zhou et al. (Zhou et al., 2023), the AlexNet achieved a slightly better AUC compared to RETFound in PD incidence prediction. The various AI architectures utilized in classification, segmentation, and PD risk assessment are illustrated in Figure 5.

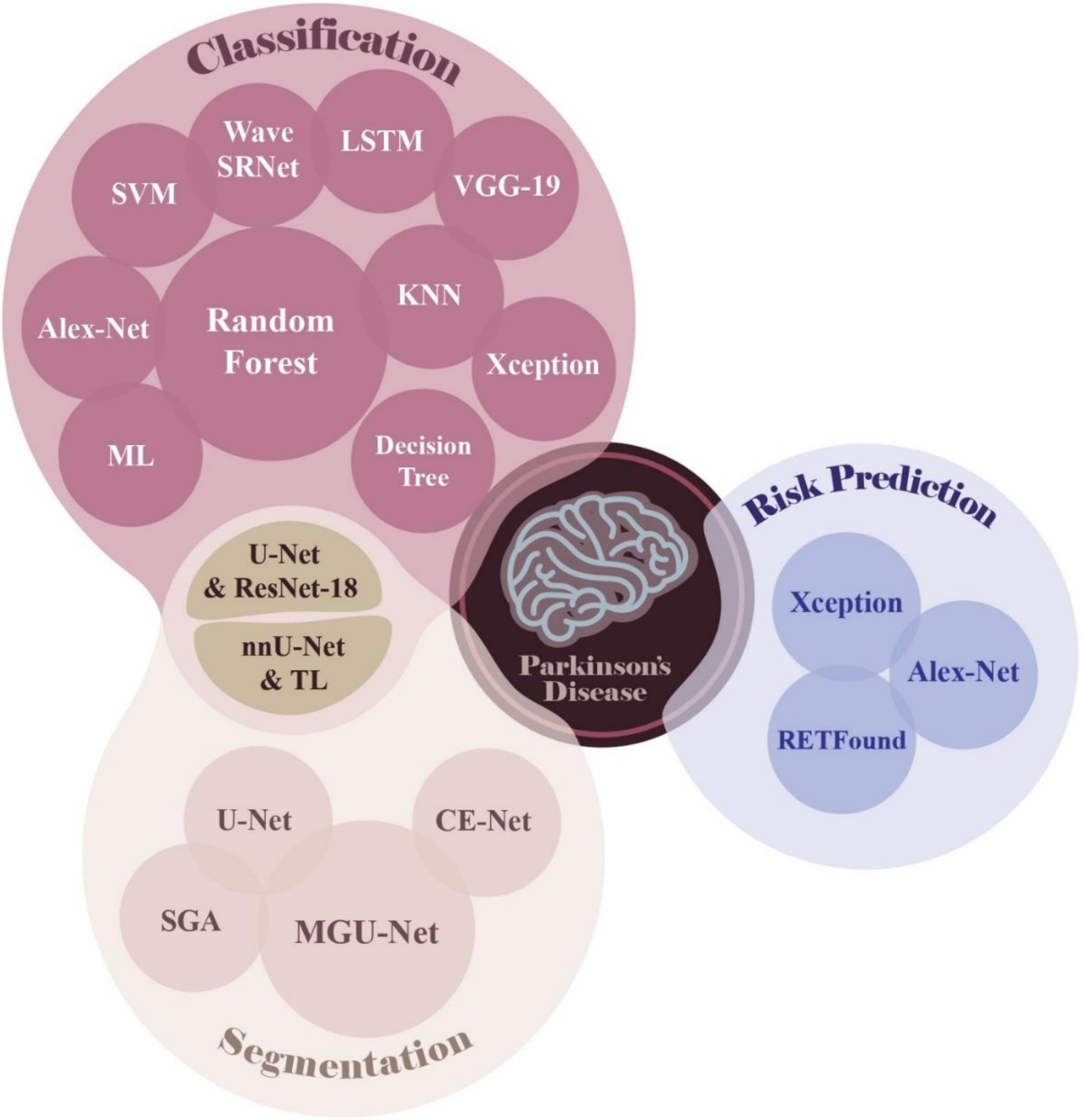

**Figure 5.** AI architectures applied for classification, segmentation, and PD risk assessment. For classification, the highest AUC was achieved with random forests, and the highest accuracy was achieved with ShAMBi-LSTM. The highest Dice score in segmentation was achieved with MGU-Net. The highest AUC risk prediction was achieved with Alex-Net

### 4.4. Other Uses

To reduce the data required to train AI models, Varghese et al. (2023) [24] used few-shot learning methods. Multiple few-shot approaches, such as TL, have been proposed; however, the technique used by Verghese et al. (2023) (Varghese et al., 2023) is called inductive logic programming (ILP). ILP logically learns facts from background knowledge and from positive and negative examples (Kononenko & Kukar, 2007). The ILP uses these inputs to infer and generate a point, and then makes decisions based on that point. Verghese et al. (2023) (Varghese et al., 2023) extracted histograms from patients' CFPs and used binning methods to enable the ILP to learn them. The ILP models were then tested for the diagnosis of multiple neurodegenerative diseases, including PD. The data presented in the figure demonstrate that ILP models, especially the Meta Inverse Entailment (MIE)–PyGol, were superior to SVM, K-nearest neighbors, random forests, and logical regression. The accuracy of Meta Inverse Entailment (MIE) – PyGol also rises as the number of positive examples increases.

## 5. Limitations and Future Direction

**Study design:** Some studies have attempted to use longitudinal data to predict future risk of PD development [52]; however, most studies employ a cross-sectional design, focusing on retinal segmentation and PD/HC classification. Levodopa treatment affects the RNFL thickness in PD (Sen et al., 2014). RNFL thickness can predict the future cognitive decline of PD (Zhang et al., 2021). These findings suggest that retinal imaging may serve as a biomarker of disease progression and treatment response, and that longitudinal studies can evaluate AI's performance in these tasks.

**Sample size and Datasets:** Only a few studies included more than 100 patients with PD (Ahn et al., 2023; Chrysou et al., 2024; Lee et al., 2023; Varghese et al., 2023; Zhou et al., 2023). It is noteworthy that the study by Hu et al. (Hu et al., 2022) used a large dataset, but, given its aim, no patients with PD were included at baseline. Only a small number of studies have used large databases such as the UK Biobank (Hu et al., 2025; Hu et al., 2022; Tran et al., 2024; Zhou et al., 2023) or iMIND (Lee et al., 2023). Other databases used in the studies were MEH-MIDAS (Zhou et al., 2023), MEH-AlzEye (Zhou et al., 2023), Heidelberg-SPECTRALIS (Álvarez-Rodríguez et al., 2024), and OCD-PD (Chen et al., 2024). Small sample sizes in studies can significantly reduce model performance and increase model uncertainty (Riley et al., 2025). The optimal sample size for each study should be calculated based on its specific setting. The type of outcome, the number of predictor variables, the prevalence of the outcome in target individuals, the desired margin of error, the desired shrinkage factor, and the signal-to-noise ratio (Riley et al., 2020). PMASAMPSIZA and PMVALSAMPSIZE are Stata models proposed for calculating the minimum required sample sizes for training and validation of models (Ensor, 2018, 2023; Riley et al., 2025). Using multiple databases can increase sample size, but this should be done carefully to reduce bias during data integration (Putrama & Martinek, 2024).

**Generative models:** Studies can also employ generative methods to artificially increase sample size. These models can detect features and generate similar data that is indistinguishable from the original ones (Reddy, 2024). The data can then be used to train an actual model. This way, studies can increase their sample sizes with high-quality images while keeping costs to a minimum. Generative adversarial networks (GANs) are among the most widely used generative AI models for data synthesis (Suthar et al., 2022). Using these methodologies is reported to increase the accuracy compared to conventional models (Gulakala et al., 2022). GANs are less frequently used in ophthalmology-related image generation than in other medical fields. However, it has been demonstrated that GANs can generate high-quality OCT images useful for educational purposes (Peng et al., 2024; Wang et al., 2021).

GANs are also used to create future CFPs for age-related macular degeneration from baseline fundus images (Pham et al., 2022). In a study by Coronado et al. (Coronado et al., 2023)GAN recreated OCTA images from CFP. This approach can effectively reduce the cost while maintaining high-quality data. These results demonstrate that GANs can be effectively applied to retinal imaging research, including OCT, OCTA, and CFP images. Although the use of generative models can improve performance, some reports indicate that they have led to a decline (Riley et al., 2025). Thus, the authors need to apply them carefully.

**Transfer learning:** We also encourage testing the proposed models with other external data sources. AI models tend to perform worse when presented with data that differs from the data on which they were fine-tuned. Most studies used ImageNet as the source domain and medical images as the target domain. ImageNet is an extensive database of mostly non-medical images (Deng et al., 2009). The similarity between the source and target domains in TL-based training approaches can enhance AI efficiency (Peng et al., 2022) and improve performance. Using PD-based databases as the source domain and fine-tuning the models on local PD data may yield more accurate models.

**Explainable AI:** Using explainable AI (XAI) models can help identify the features the AI model uses for its tasks. Only a few studies have used this method (Hu et al., 2025; Tran et al., 2024). However, the study by Hu et al. (Hu et al., 2025) applied XAI to patients with PD. Using XAI models can enable human operators to override AI results (Ali et al., 2023). Large language models (LLMs), such as ChatGPT, can also help summarize and organize the findings of other AI models and prepare an interpretable report (Parillo et al., 2024). In specific situations, LLMs have also demonstrated potential to generate plausible answers when provided with sufficient clinical data (Gilson et al., 2023). Using XAI and LLMs can help clinicians better understand the AI`s reasoning and make models more user-friendly. However, these models should be used with extreme caution, as there are reports of misdiagnoses, incorrect reports, and errors (de Bruijn et al., 2022; Parillo et al., 2024). Future studies can integrate XAI and LLMs into the current proposed methodologies for retinal imaging in PD and investigate their performance. Figure 6 summarizes the proposed future directions.

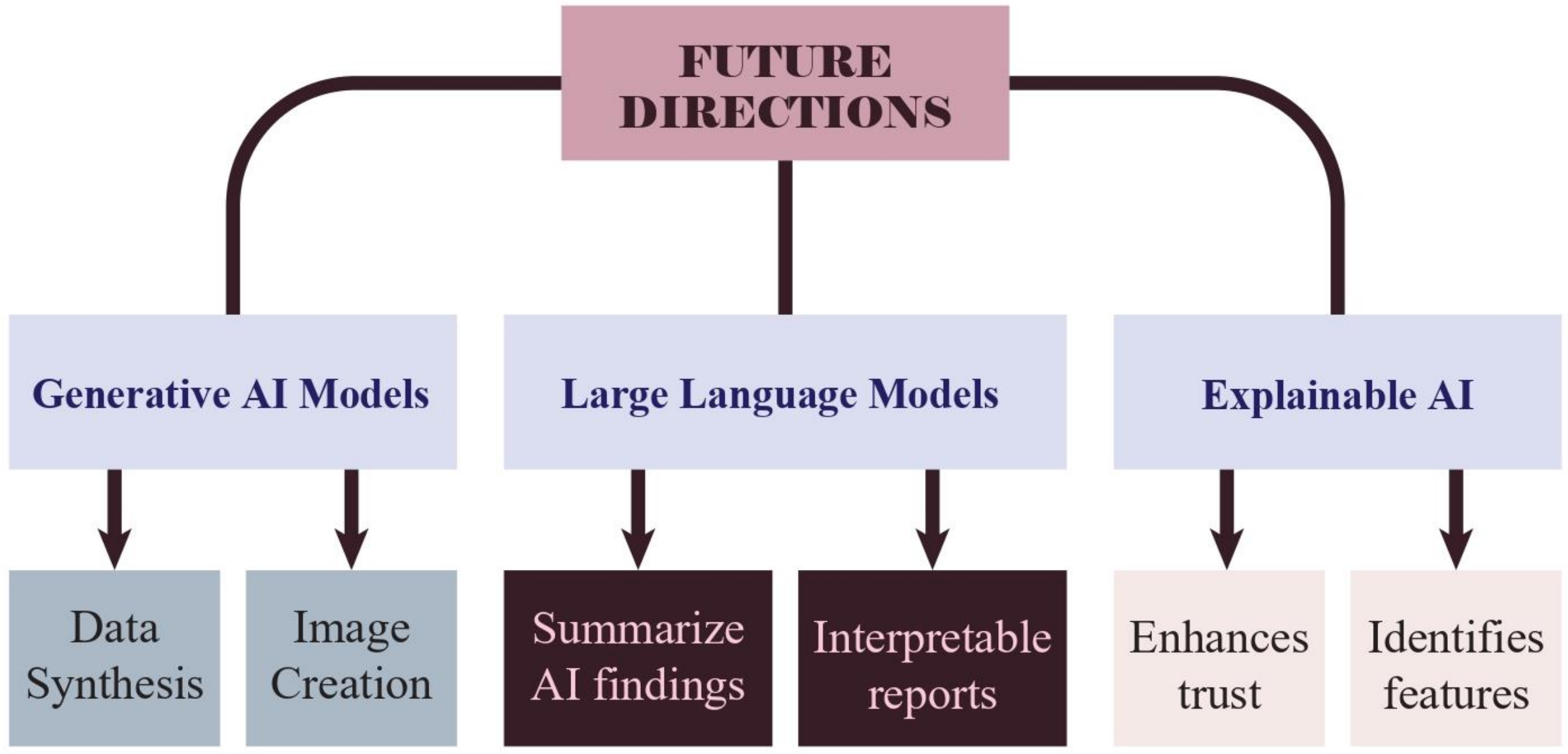

**Figure 6.** Overview of the proposed future directions in AI-assisted retinal PD research.

## 6. Conclusion

In this review, we have summarized proposed AI models for detecting PD, segmenting the retina in PD, enhancing image quality, predicting PD severity, and forecasting PD progression. Retinal thickness, particularly the RNFL and GCL, is reduced in patients with PD. The former changes are observed early in PD and can serve as valuable biomarkers for diagnosis. AI can segment retinal layers from OCT images or use texture analysis of OCTA and CFP to predict PD from HC and other neurodegenerative diseases, such as AD, WMH, and multiple sclerosis, as well as other diseases with Parkinsonian features. The performance of AI models depends on many factors, such as sample size, database balance, use of TL, AI architecture, imaging technology, image quality, task complexity, and the number of classification classes. AI shows promising results in PD-related functions. However, the use of TL, XAI, and generative models can increase the performance. Future research can elucidate the role of retinal imaging in PD progression and prognostication using AI models trained on larger longitudinal datasets.

**Abbreviations**

**AD:** Alzheimer`s Disease, **AI:** Artificial intelligence, **AMD**: Age related macular degeneration, **AUC:** area under the curve, **CE-Net:** Context encoder network, **CFP:** Color Fundus Photography, **CNN:** Convolutional Neural Network, **CNS:** central nervous system, **DL:** Deep learning, **GANs:** Generative adversarial networks, **GCC:** Ganglion cell complex, **GCL:** Ganglion Cell Layer, **GCL-IPL:** Ganglion cell inner plexiform layer, **HC:** Healthy Control, **ILP:** inductive logic programming, **IPL:** inner plexiform layer, **KNN:** K-Nearest Neighbor, **LDA:** linear discriminant analysis, **LLMs:** Large language models, **LR:** Logistic Regression, **MGU-Net:** multi-scale GCN-assisted U-shape network, **MIE:** Meta Inverse Entailment, **ML:** Machine learning, **MRI:** Magnetic resonance imaging, **MS**: Multiple sclerosis, **OCT:** Optic Coherence Tomography, **OCTA:** Optic Coherence Tomography Angiography, **OPL:** outer plexiform layer, **PD:** Parkinson`s disease, **POAG:** primary open-angle glaucoma, **RF:** Random Forest, **RNFL:** Retinal nerve fiber layer, **RPE:** retinal pigmented epithelium, **SGA**: selective-guided adversarial network, **ShAMBi-LSTM:** Shuffle Attention Mechanism depends on Bi-directional Long Short-Term Memory,

**SVM:** Support Vector Machine, **UPDRS:** Unified Parkinson's Disease Rating Scale, **UWF:** Ultra-Wide Field, **WMH:** White Matter Hyperintensity, **XAI:** explainable AI

## Declaration

**Ethics approval and consent to participate:**

This study was conducted under the principles outlined in the Declaration of Helsinki.

**Consent for publication**

N/A

**Availability of data**

N/A

**Competing interests**

The authors declare that there is no conflict of interest.

**Funding Sources**

None.

**CRedit Statement**

**Conceptualization:** AJ **Data curation:** HA, MA, PK, and AJ **Investigation:** AJ, HA, PK, and MA **Supervision:** RA, RT, RAS, and AJz **Visualization:** NS **Writing – original draft:** AJ, HA, MA, NS, and RT **Writing – review & editing:** RT, RAS, AJ, and RA **Project administration:** AJ

It should be noted that AJ and HA are co-first authors, and AJ and RAS are co-corresponding authors of this manuscript.

**Acknowledgments**

None.